\documentclass[a4paper, 12pt, twoside]{article}

\usepackage{graphicx}
\usepackage{float}
\usepackage{fancyhdr}

\title{OPTICAL COUNTERPARTS OF UNDETERMINED TYPE $\gamma$-RAY ACTIVE GALACTIC NUCLEI WITH BLAZAR-LIKE SPECTRAL ENERGY DISTRIBUTIONS}

\author{Giovanni La Mura$^{1,}$\footnote{Corresponding author: G. La Mura, {\it e-mail}: giovanni.lamura@unipd.it}, Graziano Chiaro$^1$, Stefano Ciroi$^1$, \\
              Piero Rafanelli$^1$, David Salvetti$^2$, on behalf of the Fermi-LAT collaboration, \\
              Marco Berton$^1$, Valentina Cracco$^1$}
              
\date{}

\newcommand\shorttitle{OPTICAL COUNTERPARTS OF UNDETERMINED TYPE $\gamma$-RAY AGNs}
\newcommand\shortauthor{La Mura et al.}

\makeatletter
\renewcommand\@biblabel[1]{}
\makeatother

\begin{document}

\newcommand{\tdeg}{$^\circ$}

\maketitle

\vspace{-1cm}
\begin{center}
$^1${\it Dipartimento di Fisica e Astronomia "G. Galilei", Universit\`a degli Studi di Padova, \\ Vicolo dell'Osservatorio 3, 35122 - Padova (Italy)} \\
$^2${\it IASF Milano - Istituto di Astrofisica Spaziale e Fisica Cosmica, \\ Via E. Bassini 15, 20133 - Milano (Italy)}
\end{center}

\begin{abstract}
During its first four years of scientific observations, the Fermi Large Area Telescope (Fermi-LAT) detected 3033 $\gamma$-ray sources above a 4$\sigma$  significance level. Although most of the extra-Galactic sources are active galactic nuclei (AGN) of the blazar class, other families of AGNs are observed too, while a still high fraction of detections ($\sim 30\%$) remains with uncertain association or classification. According to the currently accepted interpretation, the AGN $\gamma$-ray emission arises from inverse Compton (IC) scattering of low energy photons by relativistic particles confined in a jet that, in the case of blazars, is oriented very close to our line of sight. Taking advantage of data from radio and X-ray wavelengths, which we expect to be produced together with $\gamma$-rays, providing a much better source localization potential, we focused our attention on a sample of $\gamma$-ray Blazar Candidates of Undetermined Type (BCUs), starting a campaign of optical spectroscopic observations. The main aims of our investigation include a census of the AGN families that contribute to $\gamma$-ray emission and a study of their redshift distribution, with the subsequent implications on the intrinsic source power. We furthermore analyze which $\gamma$-ray properties can better constrain the nature of the source, thus helping in the study of objects not yet associated with a reliable low frequency counterpart. In this communication we report on the instruments and techniques used to identify the optical counterparts of $\gamma$-ray sources, we give an overview on the status of our work, and we discuss the implications of a large scale study of $\gamma$-ray emitting AGNs.
\end{abstract}


\section{Introduction}
The observation of the sky with space-born instruments, equipped with detectors working at those electro-magnetic frequencies that cannot be accessed from the ground, revealed the existence of several classes of high energy radiation sources. With their location in distant galaxies, Active Galactic Nuclei (AGNs, in brief) turned out to be the most powerful non transient sources of such radiation. Although AGNs appear with many different properties, they share an extreme intrinsic luminosity, ranging between $10^{41}\, {\rm erg\, s^{-1}}$ and $10^{46}\, {\rm erg\, s^{-1}}$, comparable to or greater than the energy output of large galaxies, but released from a region that is smaller than 1~pc in radius. To explain this property we assume that large amounts of matter, in the order of some solar masses per year, are conveyed to the nuclear regions of active galaxies, where they are accreted by a Super Massive Black Hole (SMBH). It is now well established that SMBHs with masses between $10^6\, {\rm M_\odot}$ and some $10^9\, {\rm M_\odot}$ reside in the nuclei of every massive galaxy (Ferrarese \& Merrit, 2000; Shankar, 2009). In addition, we know that accretion of fuel into their gravitational field leads to the conversion of gravitational binding energy into radiation with very high efficiency (Blandford \& Znajek, 1977; Shields, 1978).

Due to the presence of relativistic plasmas and strong magnetic fields in the vicinity of the black hole's accretion flow, the spectrum of the emitted radiation results from a combination of thermal and non-thermal components that cover several orders of magnitude in frequency, sometimes extending from radio wavelengths all the way up to $\gamma$-ray energies. The Fermi Large Area Telescope (Fermi-LAT; see Atwood et al., 2009) gave new life to the study of the $\gamma$-ray sky, producing, after 4 years of scientific observations, a map of $\gamma$-ray detections with unprecedented resolution and sensitivity in the energy range between 100~MeV and 300~GeV (Acero et al., 2015). Thanks to this result, a large number of $\gamma$-ray sources can now be associated with lower energy counterparts. In the extra-Galactic environment, it turned out that the most commonly detected objects are AGNs belonging to the blazar class. Blazars are extremely variable, highly polarized, radio-loud sources, dominated by power-law continuum spectra of type $F_\nu \propto \nu^{-\alpha}$, with a typical radio spectral slope $\alpha \leq 0.5$. Their properties are the consequence of the relativistic beaming of the synchrotron radiation produced by a jet that is collimated and accelerated close to our line of sight (Blandford \& K\"onigl, 1979). They are classically separated into BL Lac objects (BLL), whose optical spectra show a nearly featureless power-law continuum, and Flat Spectrum Radio Quasars (FSRQs), which, instead, are characterized by strong emission lines.

In addition to the blazars that dominate the extra-Galactic $\gamma$-ray population, other types of AGNs, together with a large number of sources without a firm classification, are detected as well. In this contribution we describe our investigation on the nature of $\gamma$-ray AGNs of undetermined type, through the observation of their optical spectra. We focus our attention on targets whose spectral energy distributions (SED) are consistent with those of blazars, although they still lack firm spectroscopic classification, and they are therefore called Blazar Candidates of Undetermined type (BCU in 3FGL terminology, Acero et al., 2015). In this report, we discuss the spectral classification of some BCUs, with respect to the associated SEDs, rather than giving a full list of observations. In the following sections, we describe the techniques used in the attempt to identify the low energy counterparts of the $\gamma$-ray sources, the details of how we collected the optical spectra and their classification criteria. Finally, we draw a sketch of the $\gamma$-ray emission in the different classes of objects that we observe, compared with the multiple wavelength properties of their SEDs.

\section{Association of $\gamma$-ray sources}
At the energies of $\gamma$-ray photons it is not possible to focus radiation through reflecting or refracting optical devices. The Fermi-LAT, instead, measures the production of $e^\pm$ pairs, through the conversion of photons in the detector (Atwood et al., 2009). The pair properties are used to reconstruct the energy and the direction of the incoming photon. The precision that can be achieved in the measurement depends mainly on the incoming photon incidence angle and on its energy. It can be roughly estimated that, for a photon of energy between 10 GeV and 100 GeV, hitting the detector at normal incidence, the 95\% containment angle is approximately 0.5\tdeg (Ackermann et al., 2012). Although the detection of multiple photons from a source can improve the performances, the contribution of nearby sources and background noise, combined with less than optimal detection conditions, results in a still significant uncertainty in the localization of the sources. Thus, in general, the task to associate a low energy counterpart to $\gamma$-ray sources is not trivial.

In the case of AGNs, the expected broad band emission provides a reliable way to better constrain the source position. Since $\gamma$-rays from AGNs arise from Inverse Compton (IC) scattering of low energy seed photons by relativistic plasma particles confined in strong magnetic fields, the $\gamma$-ray production occurs together with synchrotron radiation. If the energy distribution of the plasma particles is extended enough to support significant $\gamma$-ray emission, we expect that powerful synchrotron radiation is produced at radio and x-ray energies, as well. Taking advantage from the angular resolution of instruments working at these frequencies, which can measure the position of radiation sources down to a few arcseconds, we are able to identify candidate counterparts to $\gamma$-ray emission by looking for coincident x-ray and radio sources within the $\gamma$-ray detection uncertainty radius (Ackermann et al., 2011; Gasparrini et al., 2012). Furthermore, the existence of a connection between $\gamma$-ray emission and lower frequency radiation implies that large flux variations, that characterize AGNs particularly in the high energies, can be observed in different frequencies. Thanks to the monitoring strategy of Fermi-LAT observations, covering the entire sky approximately every 3~hr, the detection of important flaring activity can be matched with follow-up observations that are able to identify the possible correlated variations of the source in other frequencies. Once the source position is determined by either of the techniques down to a few arcseconds, it is possible to search for the optical counterpart and to obtain its spectrum.

\section{Optical observations}
The latest catalog of Fermi-LAT detected AGNs (3LAC, see Ackermann et al., 2015) includes blazars, radio galaxies, steep spectrum radio quasars (SSRQs), Seyfert galaxies and Narrow Line Seyfert 1 galaxies. In addition to the classified sources, however, a number of undetermined type objects, generally called AGNs or BCUs, still exists. Focusing our attention on these unclassified objects, we carried out a spectroscopic study that combines publicly available spectra, like those extracted from the latest data release of the Sloan Digital Sky Survey (SDSS-DR 12; Alam et al., 2015) and of the 6dF Galaxy Redshift Survey (6dFGRS-DR 3; Jones et al., 2004, 2009), together with new observations performed at the 1.22m and the 1.82m telescopes of the Asiago Astrophysical Observatory (Ciroi et al., 2014).

Fig. 1 illustrates examples of spectra obtained in this study. The main purpose of this optical spectroscopic analysis is to identify the AGN class for the source associated to the $\gamma$-ray emission and to determine its redshift, through the identification of known emission or absorption lines. In general, the detection of emission lines ensures the most reliable measurement of the source redshift, because they are easier to detect and they are originated very close to central SMBH. Absorption lines, on the contrary, are produced either in the host galaxy or along the light path towards us. They are much more difficult to detect and can only be seen if the AGN is not overwhelmingly dominant. It follows that the presence of emission lines characterizes AGNs with radiatively efficient accretion activity, therefore surrounded by nuclear and circum-nuclear regions of ionized gas, like quasars and Seyfert galaxies. BLLs, on the contrary, just give raise to a featureless power-law continuum, where some absorption features can show up, if the jet power is relatively low with respect to the host galaxy luminosity, or when some intervening medium happens to lie along the line of sight to the source.

\begin{table}[t]
\caption{List of sources included in this report. The table provides the $\gamma$-ray source identifier, the associated counterpart name, the counterpart position (J2000.0), the object redshift and its spectroscopic classification.}
\begin{footnotesize}
\begin{center}
\begin{tabular}{llcccr}
\hline
\hline
Id. & Counterpart name & R. A. & Dec. & {\it z} & Class \\
\hline
3FGL J$0134.5+2638$ & 1RXS J$013427.2+263846$ & $01:34:28.3$ & $+26:38:45.0$ & $\geq$0.108 & BL Lac \\
3FGL J$0339.2-1738$ & PKS $0336-177$ & $03:39:13.7$ & $-17:36:00.6$ & 0.065 & Elliptical \\
3FGL J$0904.3+4240$ & S4 $0900+42$ & $09:04:17.1$ & $+42:37:59.0$ & 1.342 & FSRQ \\
3FGL J$1031.0+7440$ & S5 $1027+74$ & $10:31:22.0$ & $+74:41:58.3$ & 0.122 & FSRQ \\
3FGL J$1315.4+1130$ & 1RXS J$131531.9+113327$ & $13:15:32.0$ & $+11:33:27.0$ & $\geq$0.730 & BL Lac \\
3FGL J$1412.0+5249$ & SBS $1410+530$ & $14:11:49.4$ & $+52:49:00.2$ & 0.076 & Elliptical \\
\hline
\end{tabular}
\end{center}
\end{footnotesize}
\end{table}
\section{Results and models}
A summary of our spectroscopic observations is presented in Table 1. According to the spectra collected in Fig. 1, our targets belong to different families of optical sources. Together with examples of classic high luminosity blazar objects, characterized by prominent emission lines (FSRQs) or dominant power-law featureless spectra (BLLs), we notice associations with some weaker sources, where the spectrum of the host galaxy, with its characteristic stellar continuum and absorption features, plays an increasing role. Looking at the spectra presented here, we note that in the high redshift regime (approximately from $z \geq 0.15$) we are able to detect $\gamma$-ray emission only from objects with powerful jets. At smaller distances, however, $\gamma$-ray sources likely arising from jets of weaker power or misaligned orientation, which do not significantly contribute to the optical spectra, can be identified as well.

In order to estimate the power of the jets, which are likely producing the observed $\gamma$-ray signal, and to compare it with the characteristics of the corresponding optical spectra, we reconstructed the SEDs associated to our targets. We used the ASI Space Data Center SED Builder tool to retrieve multiple frequency data points and to interpolate them with jet radiation models.\footnote{The SED Builder tool is available at \texttt{http://tools.asdc.asi.it/SED/}} We took into account archival data available from the literature, including in particular observations performed within extensive sky survey programs, to cover as an extended frequency range as possible. The results of this selection are plotted in Fig. 2. The broad band electro-magnetic emission is generally consistent with the classic two-hump blazar SED, very likely connected with jet activity. There are, however, some objects where an IR to optical radiation excess clearly occurs. The presence of such an excess is made particularly evident by comparing the observations with simple Synchrotron Self-Compton (SSC) models, which we used to interpolate the data. Due to the possibility that external radiation fields may give a relevant contribution to the IC effect (a scenario that is referred to as External Compton, or EC, and is very likely more appropriate, especially in the case of FSRQs), the SSC approach might be an oversimplified interpretation of the SEDs. However, the non simultaneous nature of the different data points collected in this work did not allow to consider more advanced models.

Table 2 reports details of the SED models applied to our targets. The residuals were calculated without taking into account the thermal components. The SSC models are able to reproduce most of the sources with acceptable residuals ($\chi^2_{red} \leq 1.5$), while the most notable exceptions are probably due to the difficulties that such models have in explaining the high energy IC tail. Intrinsic source variability at the different observation epochs is also expected to affect the scatter. In general, we observe that jets of large scale and power are required to interpolate classic BLL and FSRQ blazars, while weaker targets are also explained by less powerful jets.

\subsection{Notes on single sources}
{\bf 3FGL J0134.5+2638}, observed with the Asiago 1.22m telescope, reveals a power law continuum spectrum with faint absorption lines. Absence of emission lines having equivalent width $EW \geq 5\,$\AA leads to a BLL classification. Its SED shows the characteristic two-hump behavior of blazars and it is appreciably well interpolated by a SSC model. \\
{\bf 3FGL J0339.2--1738} is associated with an elliptical galaxy, with spectrum available from 6dFGRS. Flux calibration of the spectrum was obtained deriving an average sensitivity curve for the 6dF instrument, based on the flux calibrated spectra of IC 5135 and UGC 842. The resulting spectrum shows the characteristic continuum and absorption lines of an old stellar population, typical of ellipticals. The associated SED is a two-hump distribution with a prominent radiation excess in the optical window. \\
{\bf 3FGL J0904.3+4240}, detected by the SDSS, shows the highest redshift in this sample, which brings the strong UV emission lines of C~IV~$\lambda$1549, C~III]~$\lambda$1909 and Mg~II~$\lambda$2798 of quasar spectra into the optical domain. The SED is characteristic of blazars, but with a dominant IC component over the Synchrotron part. \\
{\bf 3FGL J1031.0+7440} was observed in Asiago, with the 1.82m telescope. It shows the prominent emission lines of a Seyfert 1 galaxy, with a full width at half the maximum ${\rm FWHM}({\rm H}\beta) = 2286 \pm 350\, {\rm km\, s^{-1}}$. In a standard $\Lambda$CDM cosmology with $H_0 = 70\, {\rm km\, s^{-1}\, Mpc^{-1}}$, $\Omega_\Lambda = 0.7$ and $\Omega_M = 0.3$, its redshift corresponds to a distance of 569.8~Mpc. From an apparent magnitude $V = 17.2$, we infer an absolute magnitude $M_V = -21.6$, placing this object on the border between faint quasar and bright Seyfert 1 activity. The blazar SED is accompanied by a small radiation excess in the optical domain, suggesting that the jet power and the thermal contribution from the central engine are comparable in this object. \\
{\bf 3FGL J1315.4+1130}, also detected by the SDSS, is characterized by the BLL power law continuum and by faint absorption lines. Identification of the strongest features as a Ca~II doublet is consistent with the presence of an absorption feature at the predicted wavelength of Mg~I. Fewer data points are available to reconstruct the SED of this source and we do not appreciate any deviation from a two component  blazar SED. \\
{\bf 3FGL J1412.0+5249} is detected by the SDSS and it shows the characteristics of an elliptical galaxy. Its counterpart is actually a giant elliptical located in a galaxy cluster. The associated SED is the most complex of this sample, featuring a strong optical excess, emitted by the bright host galaxy, and a high energy IC component that is hardly reproduced by SSC models. \\

\begin{table}[t]
\caption{SSC model parameters. The table columns report, respectively, the 3FGL source name, the electron energy distribution power-law index before break $\alpha_{el}^{(1)}$, the electron distribution index after break $\alpha_{el}^{(2)}$, the logarithm of the break energy (in units of $m_e c^2$), the magnetic field $B$ (expressed in Gauss), the Doppler factor $\delta$, the jet radius expressed in parsec, and the reduced residuals.}
\begin{footnotesize}
\begin{center}
\begin{tabular}{cccccccc}
\hline
\hline
Id. & $\alpha_{el}^{(1)}$ & $\alpha_{el}^{(2)}$ & $\log E_{break}$ & $B$ & $\delta$ & $R_{jet}$ & ${\chi_{red}^2}^{\rm a}$ \\
\hline
3FGL J$0134.5+2638$ & 1.5 & 4.7 & 4.0 & 1.00 & 10 & 0.001 & 1.180 \\
3FGL J$0339.2-1738$ & 1.5 & 5.0 & 4.0 & 0.75 & 15 & 0.001 & 1.097 \\
3FGL J$0904.3+4240$ & 2.3 & 3.6 & 3.5 & 0.05 & 30 & 0.003 & 1.455 \\
3FGL J$1031.0+7440$ & 1.8 & 5.0 & 4.1 & 1.00 & 15 & 0.001 & 1.339 \\
3FGL J$1315.4+1130$ & 1.6 & 4.8 & 4.8 & 0.60 & 20 & 0.002 & 1.732 \\
3FGL J$1412.0+5249$ & 1.5 & 4.7 & 4.0 & 1.00 & 10 & 0.001 & 1.889 \\
\hline
\end{tabular}
\end{center}
$^{\rm a}$ Residuals of SSC models are computed without taking into account the thermal excess data points.
\end{footnotesize}
\end{table}
\section{Conclusions}
In this study we presented the optical spectra of six targets that have been associated to $\gamma$-ray sources of still undetermined type in 3LAC. In some cases, the spectra of this type of objects can still be obtained from large spectroscopic surveys, such as the SDSS, above all, but, in others, we need specifically planned observations. The increase of sensitivity and resolution achieved by the Fermi-LAT at $\gamma$-ray energies is now providing better opportunities to identify candidate counterparts to extra-Galactic $\gamma$-ray emission and, therefore, to improve the selection of targets.

With an improved ability to detect $\gamma$-rays from faint sources we can now investigate the occurrence of nuclear activity on different power scales. The detection of faint blazar-like activity in low luminosity AGNs or even apparently normal galaxies opens a new window on the demographics of $\gamma$-ray sources, as well as on the mechanisms that contribute to black hole growth and jet formation. The possibility that such objects may represent an important contribution to the $\gamma$-ray radiation of undetermined origin deserves further investigation. Searching for radiation from hidden AGNs is a fundamental science case for instruments designed to observe high energy photons and other hints of jet activity, such as light polarization. Therefore, we plan to further investigate the spectroscopic properties of candidate counterparts to $\gamma$-ray emission, taking possibly into account light polarization studies as well, through an extensive observational campaign designed for middle class telescopes.

\section*{Acknowledgements}
The \textit{Fermi}-LAT Collaboration acknowledges support for LAT development, operation and data analysis from NASA and DOE (United States), CEA/Irfu and IN2P3/CNRS (France), ASI and INFN (Italy), MEXT, KEK, and JAXA (Japan), and the K.A.~Wallenberg Foundation, the Swedish Research Council and the National Space Board (Sweden). Science analysis support in the operations phase from INAF (Italy) and CNES (France) is also gratefully acknowledged.

This work is based on observations collected at Copernico (or/and Schmidt) telescope(s) (Asiago, Italy) of the INAF - Osservatorio Astronomico di Padova. Part of this work is based on archival data, software or online services provided by the ASI SCIENCE DATA CENTER (ASDC).

\clearpage

\begin{figure}[t]
\begin{center}
\includegraphics[width=0.45 \textwidth]{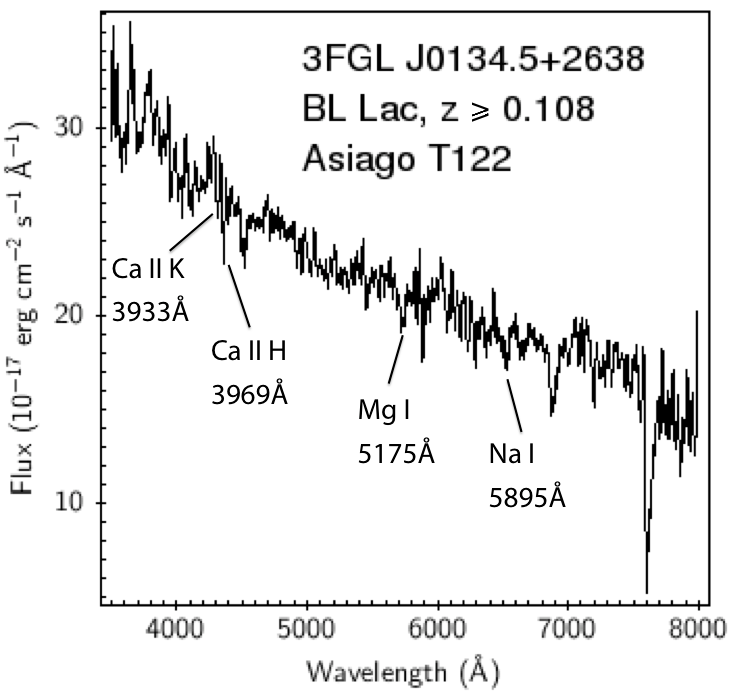} 
\includegraphics[width=0.44 \textwidth]{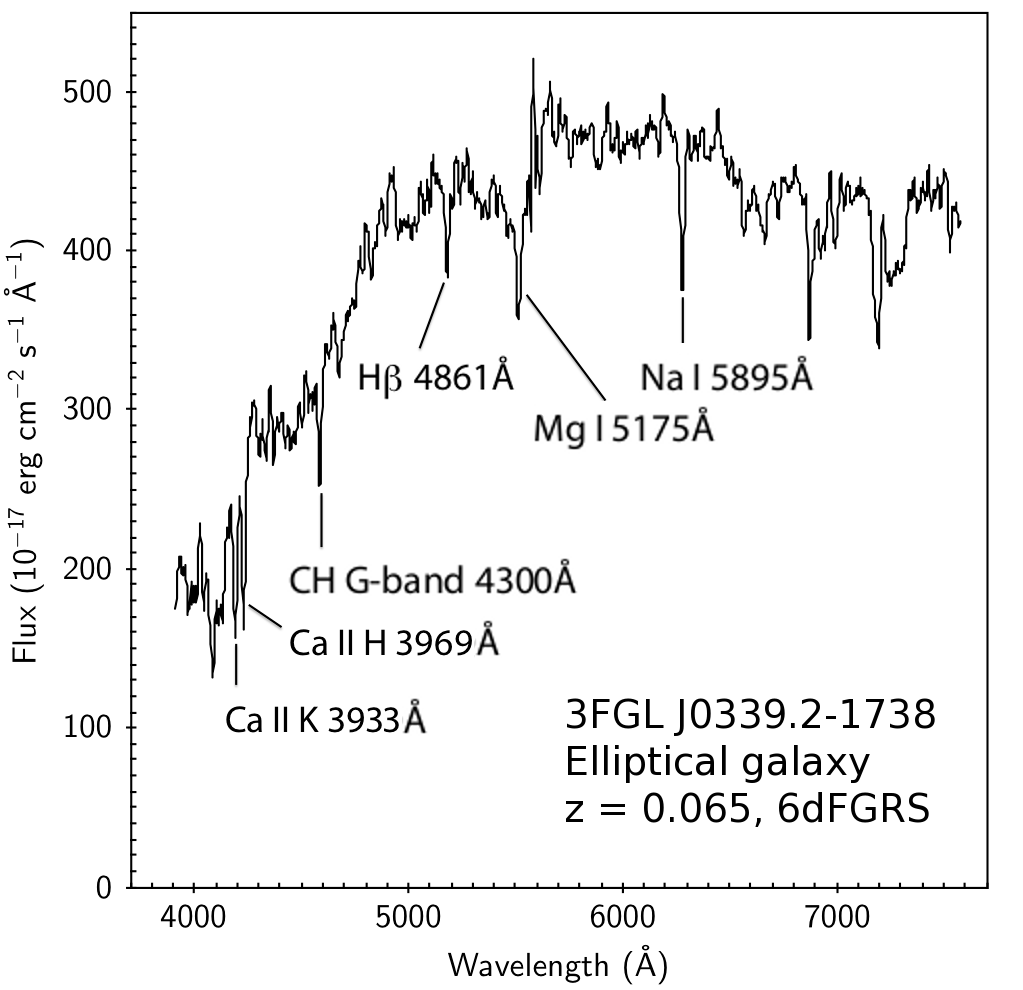} 
\includegraphics[width=0.45 \textwidth]{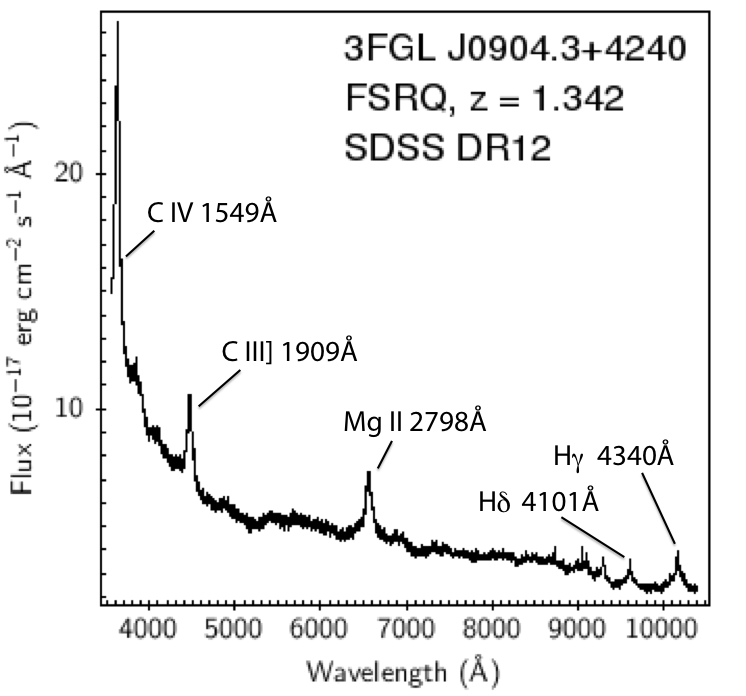} 
\includegraphics[width=0.45 \textwidth]{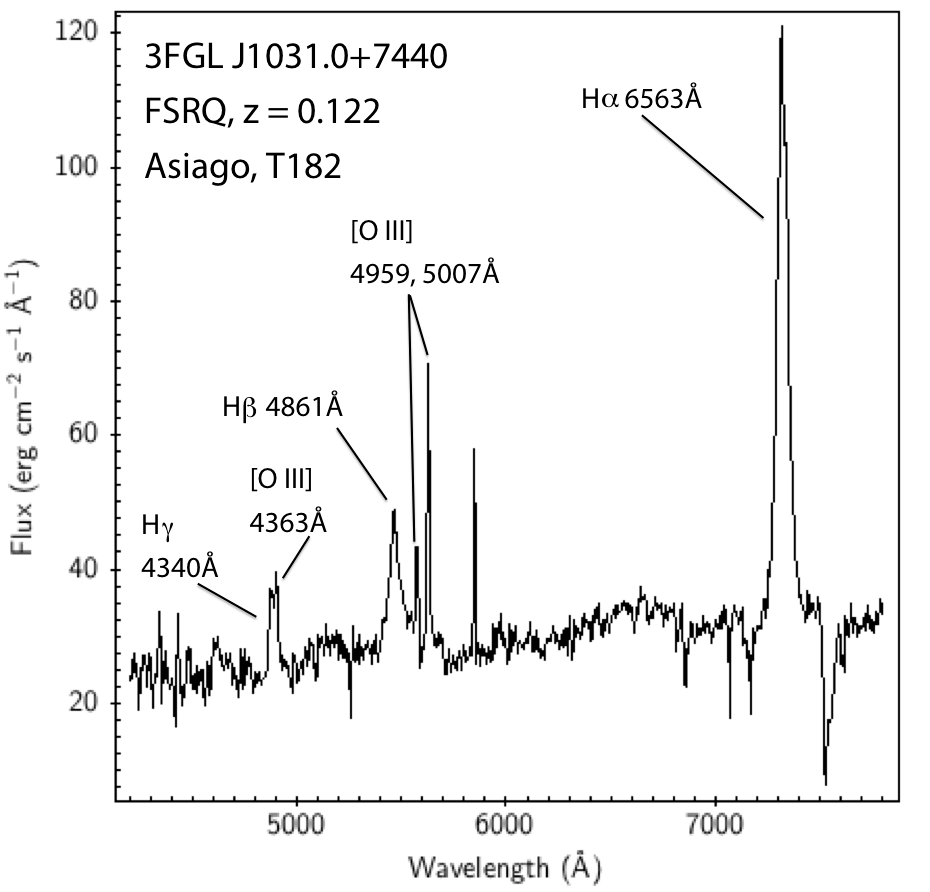} 
\includegraphics[width=0.45 \textwidth]{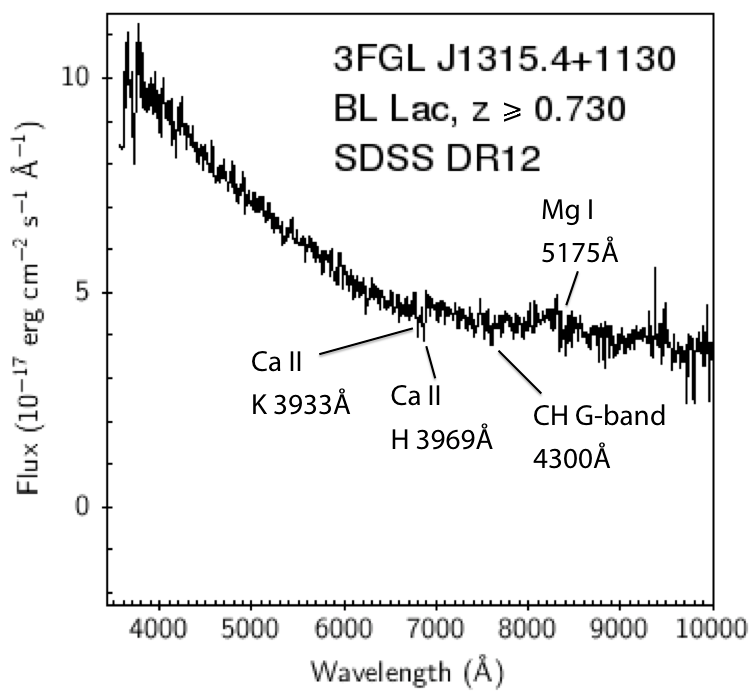} 
\includegraphics[width=0.44 \textwidth]{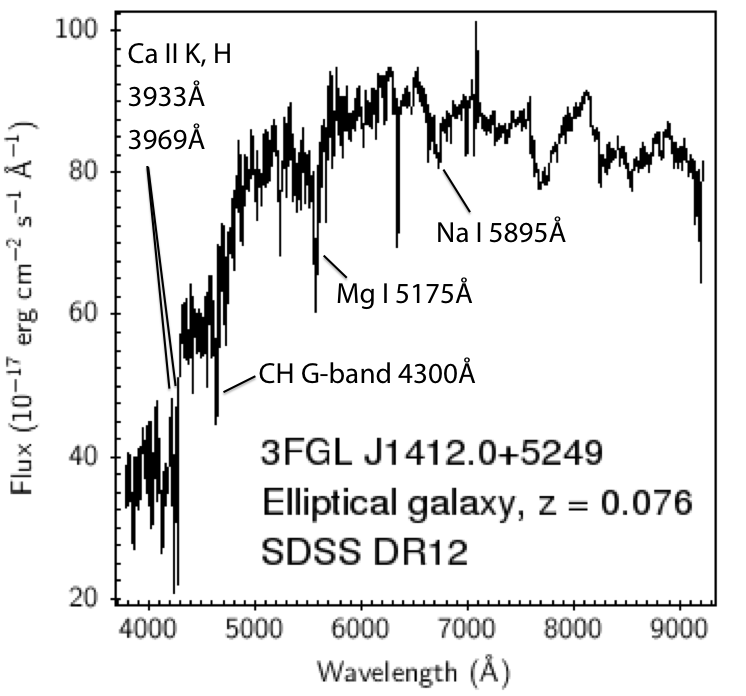} 
\end{center}
\caption{Examples of optical spectra obtained from 6 BCUs. For each object we report the associated $\gamma$-ray source, the classification, the redshift, the origin of spectroscopic data and markers for the detected spectral features. Broad emission lines are detected in objects where the central engine power is at least comparable to the jet power and they allow for firm redshift determinations. Absorption lines, on the other hand, can both arise in the host galaxy or in intervening material, so that they only place lower limits to the actual source redshift.}
\end{figure}

\begin{figure}[t]
\begin{center}
\includegraphics[width = 0.48 \textwidth]{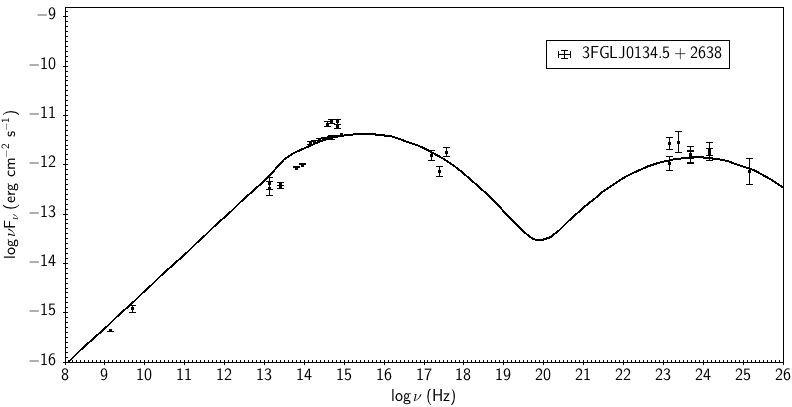}
\includegraphics[width = 0.48 \textwidth]{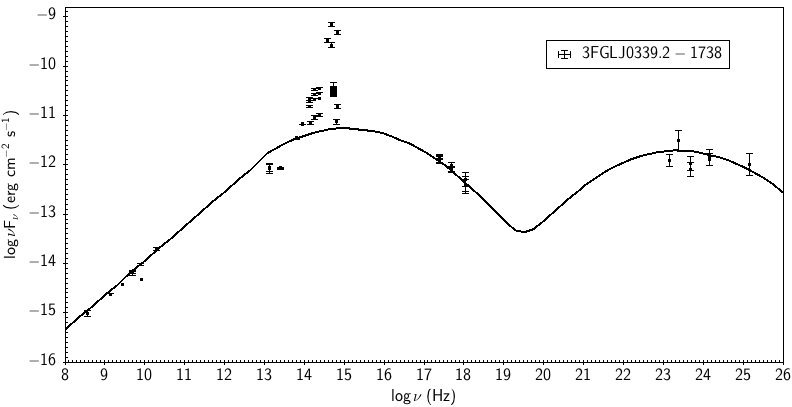}
\includegraphics[width = 0.48 \textwidth]{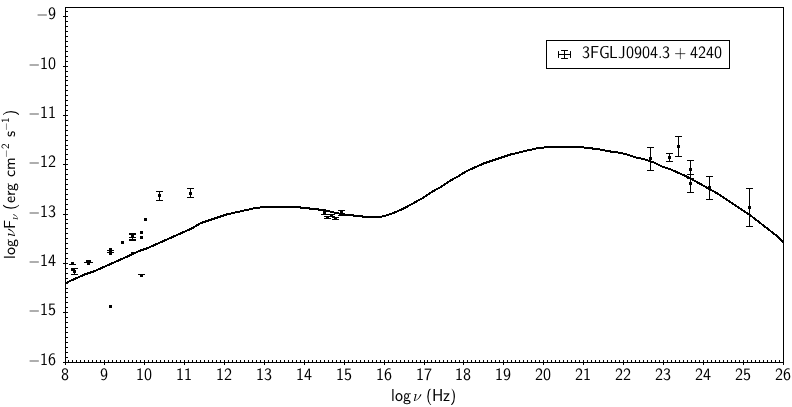}
\includegraphics[width = 0.48 \textwidth]{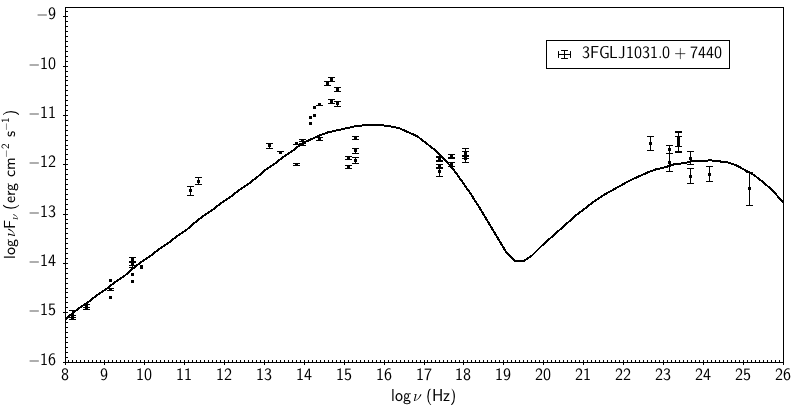}
\includegraphics[width = 0.48 \textwidth]{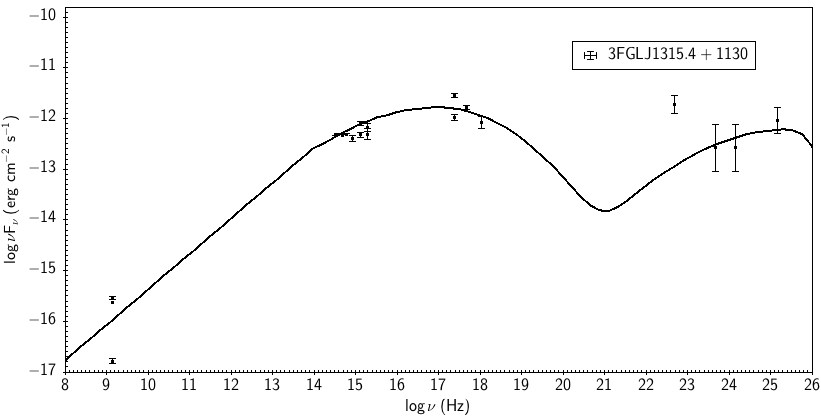}
\includegraphics[width = 0.48 \textwidth]{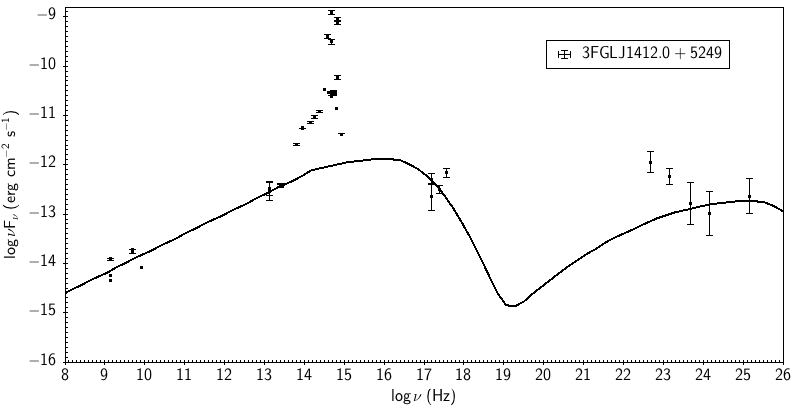}
\end{center}
\caption{The spectral energy distributions of the sources presented in this study. Every panel illustrates the multiple-frequency SED associated to the $\gamma$-ray source identified by the top label. The two-hump blazar SED is reproduced by means of SSC models, here represented with the continuous lines. It can be appreciated that thermal excesses above the SSC models are relevant in elliptical galaxies, faint in low luminosity FSRQ / Sey 1, while they are not detected elsewhere.}
\end{figure}

\end{document}